\documentclass[10pt,conference]{IEEEtran}
\IEEEoverridecommandlockouts
\usepackage{cite}
\usepackage{amsmath,amssymb,amsfonts}
\usepackage{algorithmic}
\usepackage{graphicx}
\usepackage{textcomp}
\usepackage{xcolor}
\def\BibTeX{{\rm B\kern-.05em{\sc i\kern-.025em b}\kern-.08em
    T\kern-.1667em\lower.7ex\hbox{E}\kern-.125emX}}
\begin{document}

\title{Toward Imitating Visual Attention of Experts \\ in Software Development Tasks
}

\author{
\IEEEauthorblockN{
Yoshiharu Ikutani\IEEEauthorrefmark{1},
Nishanth Koganti\IEEEauthorrefmark{1},
Hideaki Hata\IEEEauthorrefmark{1},
Takatomi Kubo\IEEEauthorrefmark{1},
Kenichi Matsumoto\IEEEauthorrefmark{1}}
\IEEEauthorblockA{\IEEEauthorrefmark{1}Nara Institute of Science and Technology, Japan\\
\{ikutani.yoshiharu.ip8, nishanth-k, hata, takatomi-k, matumoto\}@is.naist.jp}
}

\maketitle
\begin{abstract}
Expert programmers' eye-movements during source code reading are valuable sources that are considered to be associated with their domain expertise.
We advocate a vision of new intelligent systems incorporating expertise of experts for software development tasks, such as issue localization, comment generation, and code generation.
We present a conceptual framework of neural autonomous agents based on imitation learning (IL), which enables agents to mimic the visual attention of an expert via his/her eye movement.
In this framework, an autonomous agent is constructed as a context-based attention model that consists of encoder/decoder network and trained with state-action sequences generated by an experts' demonstration.
Challenges to implement an IL-based autonomous agent specialized for software development task are discussed in this paper.
\end{abstract}



\section{Introduction}
Human eye movement is an informative bio-marker that indicates domain-expertise in software development tasks, such as program comprehension~\cite{crosby2002roles, busjahn2015eye} and review~\cite{uwano2006analyzing}.
In the last three decades we have gained a lot of insight by knowing where a programmer is allocating visual attention, which can be inferred from eye movement data~\cite{sharafi2015systematic}.
A next step will be injecting these insights into an autonomous agent to efficiently perform such software development tasks.

Many have been investigating how to build an agent which can automatically perform a software development task, e.g. bug fix~\cite{gupta2017deepfix}, semantic parsing~\cite{yin2017syntactic}, patch generation~\cite{2018arXiv181207170H}, and code generation~\cite{balog2016deepcoder}.
In most cases they used only feature representations based on textual characteristics, but a few additionally utilized human gaze-fixation data~\cite{rodeghero2014improving}.
We have already known that programmers use attention strategies to save time for program comprehension and maintenance~\cite{roehm2012professional}.
For example, expert programmers tend to automatically concentrate their attention onto informative parts of a program~\cite{crosby2002roles} and skim only the relevant keywords in source code~\cite{starke2009searching}.
Incorporating gaze-fixation data allows autonomous agents to learn attention strategies that are hard to learn solely from textual characteristics.

Imitation leaning (IL) is an emerging paradigm where autonomous agents learn from human demonstration to perform complex task~\cite{argall2009survey}. 
IL views a human demonstration as an exemplar of prior knowledge in their working system and leverages a set of human demonstrations to work on tasks whose reward functions are hard to be defined {\sl a priori}.
The paradigm has been successfully applied to several applications such as autonomous driving~\cite{zhang2016query} and an initialization for reinforcement leaning in robotics~\cite{kober2013reinforcement}.
In such applications, eye movement is a major representation that can bridge human physical demonstrations and training of virtual agents~\cite{kogantiVirtual2018}.

\begin{figure*}[t]
    \centering
    \includegraphics[width=0.8\textwidth]{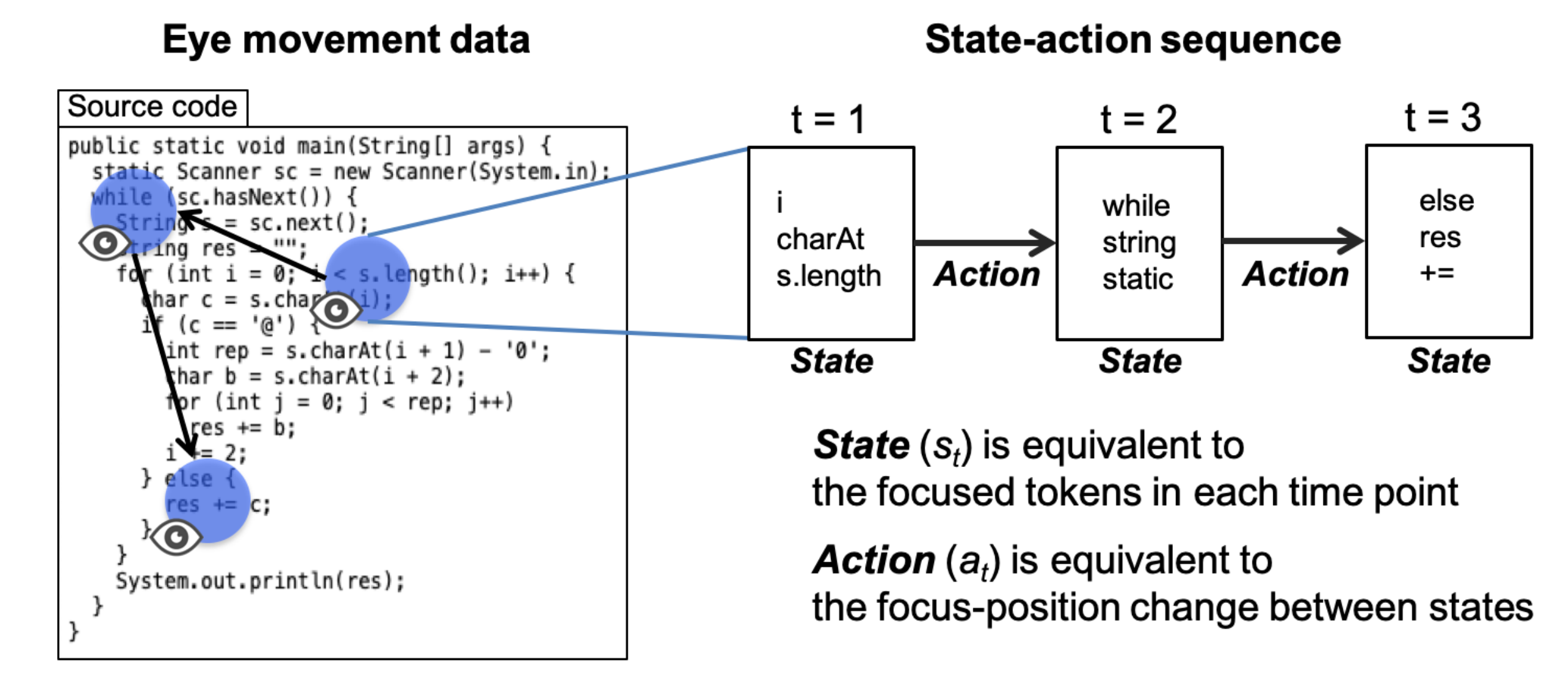}
    \caption{Programmers' eye movement as a demonstration for a state-action sequence.}
    \label{fig:seq}
\end{figure*}

We present a framework of neural autonomous agents that can learn how to view source code for a specific software development task from the demonstrations by expert programmers.
In our framework, the agent is represented by a context-based attention model that consists of encoder/decoder network~\cite{vinyals2015pointer}.
We regard programmers' gaze-fixation as a state-action sequence that represents how a programmer addressed the output that he/she finally made.
Figure~\ref{fig:seq} shows an example of state-action sequence inferred from gaze-fixation data.
The state-action sequences are created for every task period and used to train the autonomous agent to imitate the visual attention of an expert.
Historically, several mathematical models have been proposed to describe dynamics of eye movement behavior\cite{engbert2005swift}\cite{reichle2006z}. 
Their main goals were to get a biologically plausible model that can well account for real data.
In comparison with the models, a primary goal of our IL framework is to maximize agents' performance on a specific task using expert eye movements as a prior knowledge.


\section{Approach}



We propose IL for training an agent to mimic the visual attention of an expert programmer toward performing several software development tasks. The agent can be trained using the Behavioral Cloning (BC) algorithm which is commonly used in robotics~\cite{zhang2018deep} and natural language processing~\cite{hu2017learning}. The code snippet or the environment for the agent can be considered as a sequence of tokens or keywords and the agent needs to focus on a particular subset of tokens that mimics an expert programmers' visual attention. Using demonstration by an expert, the agent can be adapted to perform specific tasks such as bug fixing or algorithm detection which can be considered as an auxiliary task of the agent.

We formulate a task in IL as a sequence of states and actions. For a software development task, the state ($s_t$) is a feature representation of the current token being attended to and the action ($a_t$) is a reference to the next token. The agent performs the task using a \emph{policy} function, commonly referring to the strategy used by the agent. The policy takes as input the current state and should output the desired action. The representations used for the state and action are crucial to ensure good performance of the network. Possible candidates for software development tasks are discussed in Section~\ref{sec:featurerep}. The policy can also provide task-relevant outputs on attending to the code snippet ($o_T$). For example, for the task of algorithm classification, the task output can be a discrete label indicating the correct algorithm class. 

\begin{figure*}[t]
    \centering
    \includegraphics[width=0.85\textwidth]{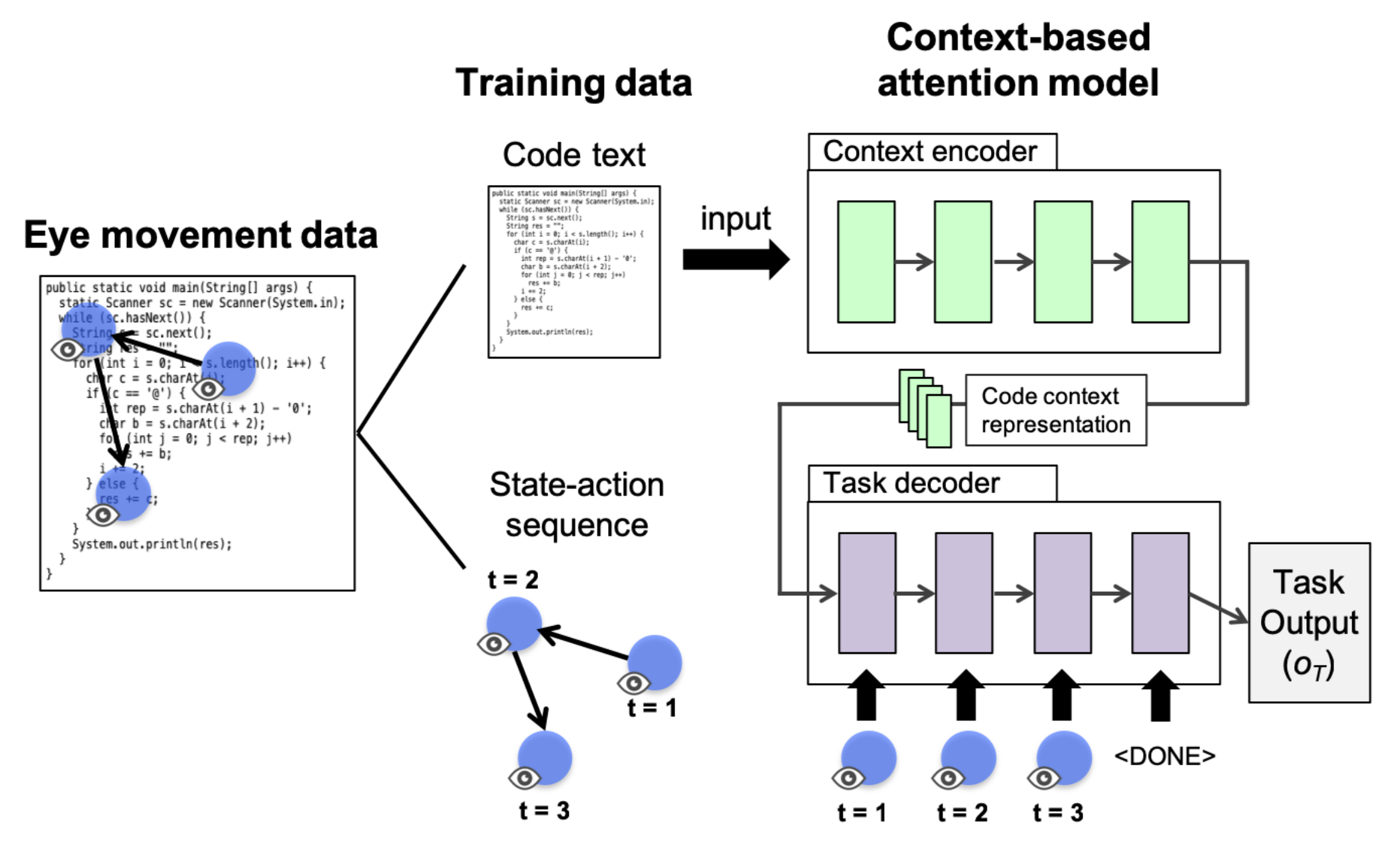}
    \caption{Overview of imitation learning agent that relies on expert's gaze data to perform source code comprehension tasks.}
    \label{fig:ilagent}
\end{figure*}

We propose to represent the agents' policy using deep neural networks, capable of learning complex feature mappings, as shown in Fig.~\ref{fig:ilagent}. As source code is a text sequence, insights from natural language processing can be used to design the agent. We propose an agent that consists of two components. The first component is a recurrent neural network (RNN) that can be used to encode the global context of the code snippet. The second component is a task-specific decoder model using an RNN that takes as input the encoded context vector at a particular token and outputs the next token to attend as the action. The task decoder also provides a task-relevant output such as a discrete label indicating algorithm class or a token index in the code snippet indicating the bug location.
Typically code snippets can be of variable lengths and requires attention over variable length sequences. To address this problem of variable size input, we propose the use of Pointer Networks~\cite{vinyals2015pointer}, which is a hard-attention mechanism capable of handling sequences of variable lengths.  

In BC, the agent is trained using two loss functions. A primary loss function is used to train the agent so as to mimic the visual attention of an expert for a particular code snippet. Depending on the task, auxiliary loss functions can be included to perform the specific task:
\begin{equation}
    \mathcal{L}_{\text{BC}}(a_t,\hat{a}_t) = w_{\text{att}} \mathcal{L}_{\text{att}}(a_t, \hat{a}_t) + w_{aux} \mathcal{L}_{\text{aux}}
\end{equation}
where $a_t,\hat{a}_t$ are the predicted action and expert action respectively. The attention loss function $\mathcal{L}_{\text{att}}$ can be a cross entropy loss with one-hot vectors representing the true and predicted tokens for attention. For the auxiliary loss function, $\mathcal{L}_{\text{aux}}$, cross-entropy loss can be used to match the algorithm class or a particular token in the code snippet sequence. The weights assigned to each loss function ($w_{\text{att}}, w_{\text{aux}}$) are assigned depending on the relative importance of the attention mechanism and task-specific output.

\section{Challenges}

\subsection{Feature Representations}
\label{sec:featurerep}

A common challenge with effective implementation of IL is the state correspondence problem where the expert and agent might have different feature representations. The feature representation should sufficiently capture the functional meaning of the code without adding any redundant information not considered by the expert. We will consider several state representations for the analysis. Source code unlike natural language is structured and so binary feature representations such as Bag-of-Words (BoW) and bag-of-n-grams~\cite{8094457,id1275} could prove effective. We will also consider word embeddings such as code2vec~\cite{alon2019code2vec} to handle long term dependencies common for source code. For the action representation, considering spatial coordinates in the image space could be ineffective as this could vary depending on irrelevant factors such as code formatting. An effective action could be location to a particular token index or to a group of tokens that represent a functional unit in the code.

\subsection{Data Efficiency for Imitation Learning}
\label{sec:dataeff}

One of the challenges of using BC is that it requires a large number of demonstrations to avoid covariate shift where the state space encountered by the agent could be drastically different from expert demonstrations. However, collecting a large dataset involving expert eye-gaze information could be prohibitive. To improve the data efficiency of IL, we will consider several strategies. A typical gaze-fixation of the human is usually on a group of tokens. This uncertainty in gaze fixation can be exploited to perform data-augmentation where a single trial can result in multiple state-action sequences by assigning different weights to the group of tokens. An alternate approach is to learn a distribution over the expert demonstrations rather than to just mimic the actions. This can be performed by using Generative Adversarial Imitation Learning (GAIL)~\cite{ho2016generative} which is based on generative adversarial training where a generator neural network is trained to model the distribution of expert demonstrations. This has been shown to provide superior performance over traditional IL in several application domains. The data-efficiency can be further improved by learning from multiple experts. However, different experts could be using different strategies and it is important to disambiguate between such latent factors. This can be performed by using techniques such as InfoGAIL~\cite{li2017infogail} which can learn to disambiguate between multiple experts through a learned discrete latent representation. 

\subsection{Possible Software Development Tasks}

The intersection of machine learning and software engineering is an emerging hot research topic~\cite{Allamanis:2018:SML:3236632.3212695}.
Our IL-based framework will open up a new way of building intelligent systems/agents.

\textbf{Issue localization}.
Fault localization is the act of identifying the locations of faults in a program. To support this time-consuming and tedious task, many machine-learning based approaches have been proposed~\cite{7390282}.
By designing a token index in the code as the task-relevant output, we can propose new systems incorporating the art of identifying bugs by experts. Preparing experimental settings for collecting experts' eye-movements is also an important challenge.
With similar approaches, we may target identifying performance issues, vulnerabilities, etc.

\textbf{Classification}.
Classification is a fundamental machine learning technique and can be applied to specific applications, such as language detection\cite{8530735}, algorithm identification.
Task-relevant outputs will be discrete labels indicating classes.

\textbf{Description generation and code generation}.
Generating the description of code or code changes based on deep neural networks is a promising active research topic, such as code summarization~\cite{Wan:2018:IAS:3238147.3238206}, commit message generation~\cite{Jiang:2017:AGC:3155562.3155583}, etc.
Generating (part of) code with deep neural networks~\cite{gupta2017deepfix,yin2017syntactic,2018arXiv181207170H,balog2016deepcoder} is another hot topic.
Although the current concept of using Pointer Networks does not work for generating sequences, the idea of incorporating experts' visual attention is also interesting for these tasks.
Designing an appropriate imitation learning framework is a desired future challenge.

\subsection{Extension with Electroencephalography as Auxiliary Input}
It may be possible to complement the system above with electroencephalography (EEG) to reflect the human cognition other than visual attention \cite{chavarriaga2010learning}.
For example, event-related potentials (ERPs) including error-related potentials (ErrPs) reflect events where a human subject feels wrong or strange. 
It is known that ERPs can reflect semantic incongruity grammatical violoations in language processing\cite{kutas1980reading, friederici2004brain}.
ERPs could be informative for programming comprehension, especially for bug fix.
To integrate gaze data and EEG data, we are considering multiple ways.
One is independent subsystem from the imitation learning,
and the other is modification of the architecture to include the EEG data as an auxiliary input.
Especially, the latter will be a tough challenge, but it will also be an attractive topic not only for software engineering but also for artificial intelligence and modelling of human cognition.  

\section{Conclusion}

A baby learns numerous things from the demonstration by parents without any lingual explanations because demonstrations can represent more than language descriptions.
So far, researchers investigated eye movements of programmers and typically converted them into human-understandable numbers and descriptions.
However, this conversion has caused considerable information loss.
In this study, we propose neural attention models trained using expert programmers' eye movements as a means of implicit learning without information loss. We presented a plausible framework to achieve this goal using IL. We also discuss several challenges that will need to be addressed to make the framework practical.
We believe that IL-based agents can fully utilize the valuable information sources with less information loss and potentially improve their performances on a variety of software development tasks.

\section*{ACKNOWLEDGMENT}
This work was supported by JSPS KAKENHI Grant Numbers JP18K18108, JP16H06569, JP16H05857 and Grant-in-Aid for JSPS Research Fellow JP18J22957.



\end{document}